# Perspective
## Sunrise for melanoma therapy – but early detection remains in the shade

Melanoma is one of the most dangerous forms of cancer. The five-year survival rate is 98% if it is detected early. However, this rate plummets to 63% for regional disease and 17% when tumors have metastasized, that is, spread to distant sites. Furthermore, the incidence of melanoma has been rising by about 3% per year, whereas the incidence of cancers that are more common is decreasing. A handful of targeted therapies have recently become available that have finally shown real promise for treatment, but for reasons that remain unclear only a fraction of patients respond long term. These drugs often increase survival by only a few months in metastatic patient groups before relapse occurs. More effective treatment may be possible if a diagnosis can be made when the tumor burden is still low. Here, an overview of the current state-of-the-art is provided along with an argument for newer technologies towards early point-of-care diagnosis of melanoma.

## Melanoma facts and figures

Skin darkens when epidermal melanocytes, a specific type of skin cell that is phenotypically different from other cells in the skin, release a dark brown pigment called melanin in the top layer of the skin in response to ultraviolet (UV) light. Nature uses melanin production as a protection mechanism against sun exposure to minimize UV light-induced damage to genetic material (DNA). In addition, skin cells have evolved complex biochemical repair systems to repair potential damage to their DNA. Unfortunately, excess damage can lead to the incorporation of oncogenic (cancer-causing) mutations that result in neoplastic transformation of benign melanocytes into malignant melanoma. Most melanoma originates within the skin, but because melanocytes also reside in other tissues, melanoma can originate in other locations such as the eye (uveal melanoma), which is exceptionally difficult to treat.

Indeed, melanoma is one of most dangerous forms of cancer in general, and skin cancer in particular. Even though melanoma only accounts for about 1% of total skin cancers, it is responsible for almost 75% of total skin cancer deaths. Major risk factors for melanoma such as unusually high numbers of moles, fair skin and/or red hair, a family history of skin cancer, and sun burns are generally well established. Unfortunately, these risk factors are often ignored as the popularity of tanning and recreational activities in sunny destinations without proper use of UV protection continue to rise. This has resulted in a disturbing trend where, in contrast to several other cancer types whose numbers are steady or decreasing (e.g., lung cancer), the incidence rate of melanoma has been increasing for over three decades, at about 2.8% per year.

In 2016 in the US alone, an estimated[1] 76,380 new cases of melanoma will be diagnosed and 10,130 patients will die from this disease. Worldwide, there are an estimated 160,000 new melanoma patients and 48,000 deaths. The lifetime risk of developing melanoma in men is currently 1:39, and in women 1:63. In 1930, this number was around 1:1,500. Melanoma is the most common form of cancer for young adults 25-29 years old and the second most common form of cancer for adolescents and young adults 15-29 years old. It is the primary cancer that causes death among

Remco A. Spanjaard[+], David Weaver[+], Shyamsunder Erramilli*, Pritiraj Mohanty*
[+]FemtoDx, 8 St. Mary's Street, Sixth Floor, Boston, MA 02215, USA
*Department of Physics, Boston University, 590 Commonwealth Avenue, Boston, MA 02215
*Center for Nanotechnology and Nanobiotechnology, 8 St. Mary's Street, Boston, MA 02215
Corresponding author (Email: mohanty@physics.bu.edu (PM))

women aged 20-25 years, while only trailing breast cancer in 30-35 year old women. These numbers indicate that melanoma has become an increasingly urgent public health issue. This is not likely going to become less of a problem in the foreseeable future, not only from a clinical perspective, but also economically because the high cost of typically failing metastatic melanoma patient therapy significant taxes an already overburdened health care system.

## Visual diagnosis: An inexact science

Focusing on skin, the most commonly used mnemonic[2] to diagnose a suspicious mole (also known as a nevus) as a potential melanoma is the so-called "ABCD". A stands for Asymmetry, because the typical shape of a mole gets lost and one half is different from the other. B stands for Border Irregularity because the edges are often uneven. C stands for Color because rather than a uniform shade of brown, a melanoma may consist of a mix of brown, black and tan areas. Finally, D stands for a Diameter over 6 mm because larger moles are more likely to be malignant than small ones. E is sometimes added for Enlarging or Evolving, indicative of a growing population of potentially cancerous cells. Warning signs other than those listed in the ABCD/E include changes in sensitivity or bleeding. When a patient with a suspicious mole sees a doctor, typically a punch biopsy is ordered and the pathologist will diagnose based on histological and immunohistochemical markers. If the diagnosis is confirmed as melanoma, the tumor will be staged according to the criteria shown in Table 1.

**Table 1.**     **Melanoma staging and five-year survival rates**

| | |
|---|---|
| **Stage 0**: | Melanoma *in situ*. In majority of cases, these tumors can be easily resected and five-year survival rates are 99.9%. |
| **Stage I / II**: | Invasive melanoma. These comprise smaller tumors (< 1 mm) with or without ulceration, or larger (< up to 2 mm) without ulceration. Five-year survival rates are 89–95%. |
| **Stage II**: | High risk invasive melanoma. A group of tumors ranging in size from 1.01- 4 mm in thickness further subdivided pending absence or presence of ulceration. Five-year survival rates are 45–79%. |
| **Stage III**: | Local or regional metastasis. This entails involvement of up to four positive lymph nodes. Five-year survival rate is 24–70% with progressively worse prognosis if more lymph nodes have presence of tumor, or if a single lymph node is positive in combination with regional or skin metastases. |
| **Stage IV**: | Distant metastasis showing widespread disseminated disease and organ involvement. Further sub-classification takes lung metastasis and lactate dehydrogenase (LDH) levels into account, with overall poor five-year survival rates of 7–19%. |

Of critical prognostic importance are the assigned "Clark level" and "Breslow's depth", which refer to the vertical rather than horizontal growth phase of the tumor. In the absence of any other reliable biomarkers, this is one of the best clinical predictors of metastasis.

Unfortunately, the "ABCD/E" technique has many problems. Multiple types of melanoma are recognized but not all of these can be detected by "ABCD/E". Furthermore, many melanomas will not display all four or five characteristics simultaneously. This poses a big challenge for the correct diagnosis of the estimated 60-70 million moles that are screened annually. In the absence of more specific non-invasive tests, about 10% of these suspicious moles are excised but only 3% are actually cancerous moles. In other words, 97% or almost all of the 6-7 million excised moles each year are unnecessarily removed.

## Emerging non-invasive technologies to diagnose suspicious skin moles

Several companies have realized that a diagnostic test that can differentiate between a mole that is malignant and one that is atypical, but otherwise benign, can be of great clinical and commercial importance. The market size for screening moles is $2-$3B per year and that of subsequent analysis of biopsies is about $400M. Four leading technologies that address this issue are listed in Table 2. Furthest regulatory development has been achieved for MelaFind and Symsys Molemate, which are the only two technologies to have received recent FDA marketing approval. MelaFind uses light to differentiate benign from malignant moles based on their specific absorption profiles while Symsys-Molemate employs intracutaneous spectrophotometry to analyze vascular composition and pigmentation. In a somewhat analogous fashion, NeviSense and Aura use the inherent differences in electrical impedance and Raman spectra, respectively, between normal and malignant nevi to correctly identify only those lesions that must be removed. All four devices are designed to be used in a point-of-care (POC) setting to support and advise rather than replace the pathologist's assessments, and thereby reduce the number of unnecessary excisions. However, it is noteworthy that even MelaFind is far from being accepted, let alone adopted by the medical establishment due to its heavy bias towards (false) positive identification of malignancy. Furthermore, the devices can only be used on moles of certain sizes and degree of pigmentation, such that the anatomical location and lesion condition (e.g., scarring, ulceration) of a mole can further restrict their usage. An alternative type of technology is being developed by DermTech based on genetic markers for melanoma. Suspicious cells are lifted from the lesion by adhesive tape and the RNA expression profile is obtained by quantitative polymerase chain reaction (qPCR) techniques and compared to a proprietary panel of melanoma-associated genes to correctly diagnose the mole. Claimed sensitivity and specificity of this technology is extremely high. However, it is not a POC technology, validation is still ongoing, and the FDA approval date is uncertain.

These technologies still require varying levels of refinement. However, it is encouraging that considerable advances have been made to aid the pathologist in reaching a correct diagnosis of skin lesions, which is currently done with < 20% accuracy. It is conceivable that at least one of these technologies will eventually become part of the standard of care with clear benefits for the payer and the patient. However, it is important to note that while these diagnostic devices prevent many unnecessary biopsies, they do not impact treatment or survival of metastatic melanoma patients.

## New therapies provide hope to metastatic melanoma patients within limits

Conventional treatments for melanoma are surgery (very successful in early stage melanoma), radiotherapy, chemotherapy (e.g., dacarbazine), and immunotherapy. The latter essentially consisted of use of interferon alfa-2B, a formulation that was improved upon in 2011 by a pegylated version (Sylatron). All of these treatments are nevertheless essentially helpless against metastatic melanoma, a situation that has remained unchanged until in 2011 the first of 6 new targeted therapies presented advanced stage disease patients with exciting benefits (Table 3). Even though, in absolute terms the effects on survival are relatively small, on the order of 3-6 months due to invariably occurring resistance and relapse of the patients, a large percentage of patients respond significantly to the drugs, something that was never before seen in metastatic melanoma on this scale. The first approved immunotherapeutic drug is Yervoy, an antibody that targets and inhibits CTLA-4, a negative regulator of cytotoxic T-cell activity that is critical for mounting an effective antitumor immune response. Using a similar strategy, two other antibodies, named Keytruda and Opdivo, block the interaction of PD-1 on T-cells with its ligands PD-L1 and PD-L2 to reactivate antitumor immunity in melanoma and other cancer types. This mechanism received "Breakthrough Therapy" status from the FDA.

**Table 2.        The most advanced non-invasive technologies for diagnosis of skin moles**

| Trade name | Technology and Advantages/Disadvantages |
|---|---|
| | **HAND HELD OPTICAL SCANNING DEVICES** |
| MELAFIND | - Developed by MELA Sciences (US), POC device uses light to image the skin through a layer of isopropyl alcohol to generate a positive or a negative result based on predefined image analysis algorithms. Sensitivity is 98.3% and specificity 10.8%. FDA approved on November 2011.<br>- Not covered by health insurance.<br>- Usefulness heavily debated due to engineered bias towards positive results, thus negating its intended advantage over traditional pathology exams. |
| NEVISENSE | - Developed by SciBase (Sweden), POC device uses electric impedance spectroscopy to determine if a mole is benign or malignant melanoma with claimed 97% sensitivity and 34% specificity.<br>- FDA approval has not yet been obtained although device was approved in Australia in October 2013. |
| AURA | - Developed by Verisante Technology (Canada), POC device uses Raman spectroscopy to measure differential vibrational modes of biomolecules between benign and melanoma skin lesions. At a sensitivity level of 90-99%, positive predicted values were between 15-30% and negative predicted values were at 98-99%.<br>- Approved by Health Canada as Class II Device in 2011, but FDA approval has not yet been obtained. |
| SIMSYS-MOLEMATE | - Developed by MedX Health (Canada), POC device uses spectrophotometric intracutaneous analysis to image and analyze vascular composition and melanin of suspicious skin lesions.<br>- Approved by FDA in September 2011, by Health Canada as Class II Device in January 2012, and can also be marketed in Europe.<br>- Wide applicability not limited to melanoma.<br>- A UK study showed that the device operates at a cost of £18 over best practice and only adds 0.01 quality-adjusted life-year per patient. Moreover, use of MoleMate resulted in more referrals and evidence suggests that best practice is still more accurate. |
| | **OTHER TECHNOLOGIES** |
| PIGMENTED LESION ASSAY | - Developed by DermTech (US), skin cells from a mole are lifted using a proprietary Adhesive Skin Biopsy Kit. Next, a qPCR-based assay analyzes RNA expression profiles of these cells to determine if they are associated with melanoma or non-melanoma. Claimed accuracy is 91%, with 91% sensitivity and 69% specificity.<br>- Complex analysis is not suitable for POC.<br>- FDA approval date is unknown. |

Distinct from antibody-based strategies are a new generation of targeted small molecules, of which Zelboraf (vemurafenib) was the first to be approved. About half of all melanomas have a mutation in the *BRAF* gene resulting in a permanently active form of BRAF called BRAF$^{V600E}$ or BRAF$^{V600K}$. BRAF is a signaling molecule that helps tumor cells proliferate. Zelboraf inhibits the mutated BRAF forms in the tumor cells and has been shown to shrink many of these tumors. Zelboraf, and similarly designed Tafinlar (dabrafenib) are now used in melanomas that test positive for the *BRAF$^{V600E}$* gene change in companion diagnostic assays. Finally, Mekinist (trametinib) targets a downstream effector of BRAF called MEK. Mekinist is approved for treatment of tumors that carry either the BRAF$^{V600E}$ or BRAF$^{V600K}$ mutation, which is also determined by a PCR-based companion diagnostic test.

**Table 3.    Recently approved therapies in (metastatic) melanoma**

| Trade name (USAN) | Approval date | Manufacturer | Drug target |
| --- | --- | --- | --- |
| Sylatron (peginterferon-alfa-2B) | March 29, 2011 | Schering | Immunomodulatory genes |
| YERVOY (ipilimumab) | March 25, 2011 | Bristol-Myers Squibb | CTLA-4 |
| KEYTRUDA (pembrolizumab) | September 4, 2014 | Merck | PD-1 |
| OPDIVO (nivolumab) | December 22, 2014 | Bristol-Myers Squibb | PD-L1 |
| ZELBORAF (vemurafenib) | August 17, 2011 | Genentech/Roche | BRAF$^{V600E}$ |
| TAFINLAR (dabrafenib) | May 29, 2013 | GlaxoSmithKline | BRAF$^{V600E}$ |
| MEKINIST (trametinib) | May 29, 2013 | GlaxoSmithKline | MEK |

There are indications that, once durable responses are achieved, patients may remain disease free. However, as promising as these results may be, there remains a huge unmet clinical need for therapeutics that can achieve longer lasting responses in the majority of patients.

Alternatively, a diagnostic test that can detect early-stage metastatic disease by measuring the presence of melanoma biomarkers or circulating melanoma cells (CMCs) in the bloodstream before the cells colonize distant organs may dramatically impact long-term survival in this patient group. It is conceivable that when the tumor burden is low, melanoma cells may be more responsive to (chemo)therapy because less time has been available for potentially resistant genetic variants to develop. This paradigm has never been tested, because technologies that can reliably detect very small numbers of circulating or metastasized melanoma cells are extremely limited in their capacities and reliable biomarkers are not currently known to exist in blood. However, new technologies are being developed, which, combined with new biomarker discoveries, may change this situation.

# Quantifying circulating tumor cells as a prognostic tool

When cancer cells detach from the primary tumor and enter the bloodstream, they can circulate through the body and invade distant organs, a process known as metastasis. Metastasis is the typical cause of death for a cancer patient. Circulating tumor cells (CTCs) have been detected in a variety of carcinomas and they appear at widely varying frequencies. The hypothesis that is supported by a number of clinical studies is that the number of CTCs is associated with reduced progression-free and overall survival. As such, a device that can quantify CTCs can be used as an early warning prognostic tool when more conventional imaging technologies are not adequately sensitive. Direct detection of CTCs is technically very demanding because their typical presence is only 1-10 CTCs per $10^6$ (1-10 parts per million) blood cells. Numerous technical strategies have been evaluated for the direct detection of CTCs, but there are only two viable technologies, as discussed below.

**PCR.** PCR is an extremely sensitive technique for detecting the presence of minute quantities of genetic material (RNA or DNA). PCR has been hypothesized to be able to detect melanoma-associated RNA as a marker for CMCs. In 1998, the University of Chicago began a trial with the NCI (NCT00004153). The goal was to determine if performing nested reverse transcriptase-PCR (RT-PCR) on a panel of five different melanoma-associated genes to increase sensitivity and specificity over single gene analysis, using lymph node samples or peripheral blood from melanoma patients, could predict relapse and disease stage. 106 patients were enrolled and followed for two years, and the study was completed in 2005. Unfortunately, the RT-PCR-based analysis suffered from inherent major technical issues and no study results were ever reported on ClinicalTrials.Gov. Additional PCR-based clinical studies have shown that the correlation of increasing numbers of CMCs and worse disease stage and prognosis is generally correct, but not perfect. Moreover, the PCR-based approaches remain technically troublesome. A promising variant of this approach was presented by Dundee University (UK) researchers, who claimed that levels of TFP12 DNA in patient blood could be a key marker to correctly predict disease progression; however, these studies still need validation in larger clinical trials.

**CellSearch System.** CellSearch System is a device marketed by Veridex (US). This technology uses antibodies that selectively adhere to a specific molecule type on the CTC surface, for example EpCam in case of epithelial tumors. Once the antibody attaches to its target, the CTCs are separated from the blood cells by directing them into different microfluidic channels. The claimed analytical sensitivity is 1 CTC/7.5 mL blood (or 1 CTC per 40 billion blood cells) with a specificity of 99.7%. The device was first developed for use in breast cancer where patients with >5 CTCs/7.5 ml blood were found to have a considerably worse outlook than those with <5 CTCs/7.5 ml. The FDA approved the CellSearch System to predict progression-free survival (PFS) and overall survival (OS) in patients with metastatic breast cancer, which was subsequently expanded on March 8, 2008 to include colorectal and prostate cancer. The CellSearch System is also available in Europe. Similar to CellSearch, the Maintrac CTC Count Test uses fluorescence-based enumeration of Circulating Epithelial Tumor Cells (CETC) through quantitative microscopic analysis of epithelial cells expressing Epcam that are bound by FITC-conjugated anti-CD45 antibodies. Unfortunately these devices have significant drawbacks. For instance, not all epithelial tumor cells express Epcam, which would confound any interpretations based on set critical number limits. A 3$^{rd}$ generation CTC-iChip is now being developed at Massachusetts General Hospital (US). It plans to address the issue by using magnetic beads and filtration to actively remove the blood cells. Furthermore, melanoma is not an epithelial cancer, therefore it requires targeting of melanoma-specific cell surface proteins. This requires extensive R&D and clinical validations of candidate biomarkers, as the long-term viability of the CellSearch System has already been questioned by some analysts.

# Surrogate biomarkers for the detection of early stage melanoma: Novel biomarker TROY

In light of the many technical issues associated with detection of CMCs, a large number of secreted or shed serum factors have been analyzed as potential surrogate diagnostic and/or prognostic biomarkers. Unfortunately, they also suffer from reliability and/or sensitivity problems that prevent their use in a clinical setting. These include today's most widely-used melanoma immunohistochemical biomarkers S100b, HMB-45 and MART-1.[3,4] Therefore, there is no gold standard or imminent technology with the capability of detecting occult metastatic melanoma in blood either directly or indirectly. However, a new candidate melanoma biomarker has recently been discovered, which, in combination with a novel ultrasensitive point-of-care detection technology, may address this unmet need.

TROY (TNFRSF19) is a type I orphan member of the tumor necrosis factor receptor superfamily (TNFRSF).[5] It is ubiquitously expressed during embryonic development where it may control cell proliferation and inhibit differentiation[6] through binding with lymphotoxin-alpha[7] or perhaps an as-yet-unidentified ligand.[8] After birth, expression is restricted to hair follicles and the brain,[9,10] where it functions as a inhibitory coreceptor of nerve cell regeneration.[11,12] However, there is increasing evidence linking disregulation of TROY expression to human disease. TROY is upregulated in multiple sclerosis and spinal cord injury, suggesting a role in central nervous system pathologies.[13] TROY mRNA is overexpressed in advanced glial tumors and its elevated expression both correlates with a poor patient outcome and increases the migration and invasion capabilities of glioblastoma cell lines.[14] TROY is expressed in all primary and metastatic melanoma cells, but not healthy melanocytes or other normal cells or tissues with the exception of sebaceous glands.[15] Thus, TROY is an aberrantly re-expressed embryonic gene in melanoma. To our knowledge, it is the only melanoma-specific cell surface biomarker. Type I membrane receptors such as TNFR1 (which binds TNFalpha) and EGFR are known to shed their extracellular domain (ECD).[16,17] The presence of melanoma-derived TROY ECD in patient serum would therefore present a unique opportunity for a diagnostic blood test as a surrogate marker for CMCs and occult metastasis. Additionally, when combined with an innovative sensor platform that performs ultrasensitive, real-time point-of-care measurements of shed TROY in ECD, such a test has the potential to change the standard of care. A possible melanoma treatment plan integrating this diagnostic technique is shown in Figure 1.

**Life-long monitoring of melanoma patients who have undergone therapy but are at risk of relapse for presence of CMCs.** Distinct from diagnosis of new patients, there are additional uses of a TROY-based diagnostic test. Cure rates for melanoma never reach 100%. Even commonly diagnosed stage I/II patients already have a 5-11% chance of relapse. Relapse probabilities become worse as staging increases, indicating that in those cases there already is widespread, but undetectable (occult) metastatic disease. These at-risk patients that are clinically free of disease could be subjected to periodic testing by our CMC-detecting device. Patients with elevated levels could be treated immediately to prevent establishment of metastatic tumors in distant organs. Finally, TROY serum levels can be utilized as a read-out for the success of treatment of established metastatic melanoma patients. If TROY levels do not decrease, indicating resistance, an alternative therapeutic approach would be selected. This would be highly beneficial to the patient because it would reduce time spent on ineffective treatments with potential associated side effects, toxicity, and morbidity, and it would expedite selection of the best available treatment for a given patient.

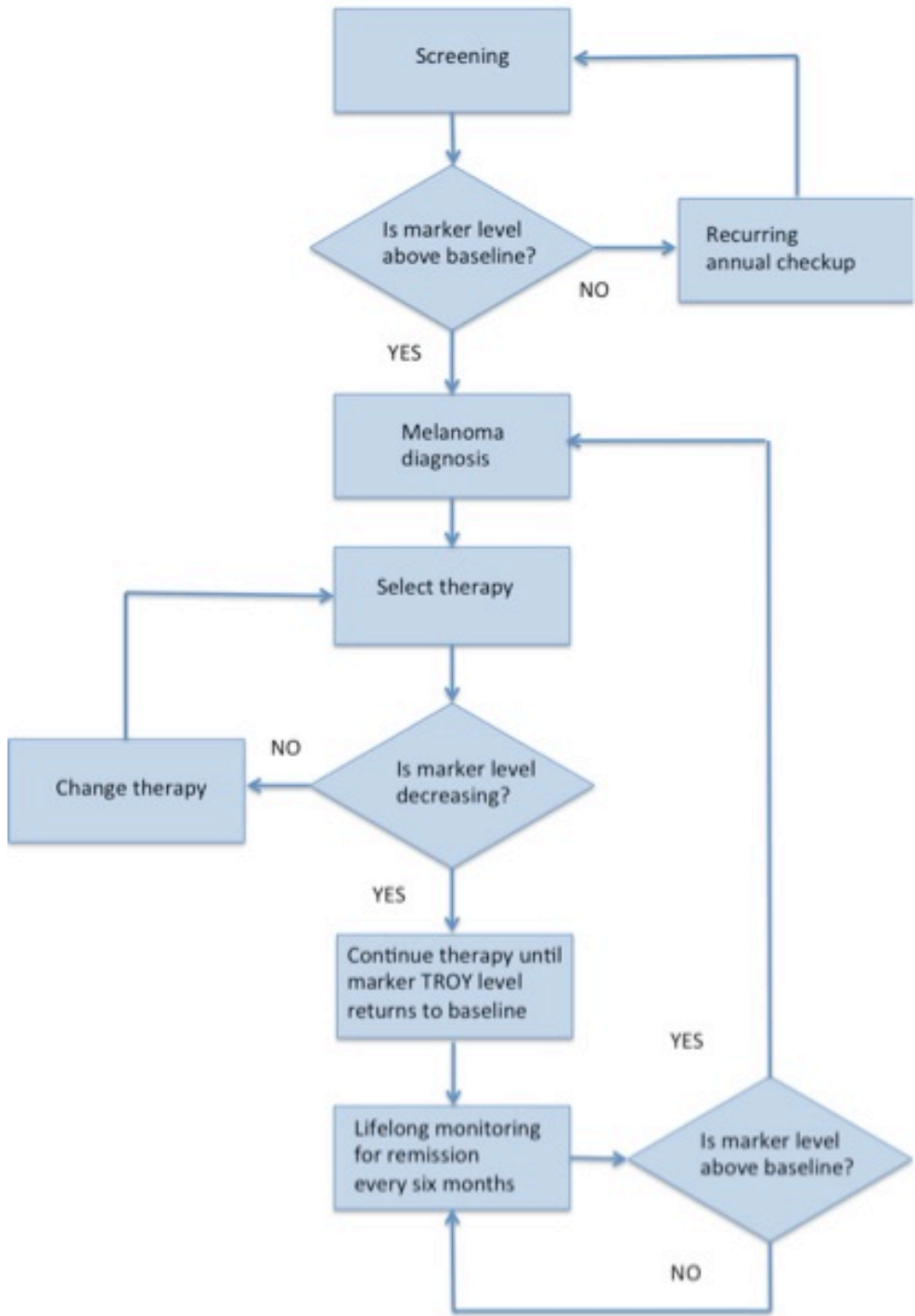

**Figure 1.** The proposed diagram demonstrates the potential use for the TROY sensor in diagnosis and prognosis of melanoma.

# Point-of-care diagnostic blood test for melanoma

A Point-Of-Care (POC) diagnostic blood test that can detect abnormally elevated levels of TROY ECD as a direct read-out for the presence of CMCs and occult distant metastasis can revolutionize diagnosis and treatment of melanoma. The primary application for such a novel device is early detection of melanoma using the proposed blood test. Because no such diagnostics test exists in the market at this time, an estimated 400,000 patients may go undiagnosed every year. Not only is early diagnosis of these patients using a POC blood test expected to positively impact survival rate, the test can also be used for monitoring of therapy responsiveness and lifelong monitoring.

There are numerous technologies capable of detecting protein markers in blood with the required sensitivity and specificity. However, these incumbent technologies can only be employed in a central laboratory setup. These methods used to measure (serum) protein levels are proteomic technologies, such as fluorescent imaging on microarrays, or they are conventional technologies like enzyme-linked immunosorbent assay (ELISA) that are not well suited for real-time, POC diagnostics.[18]

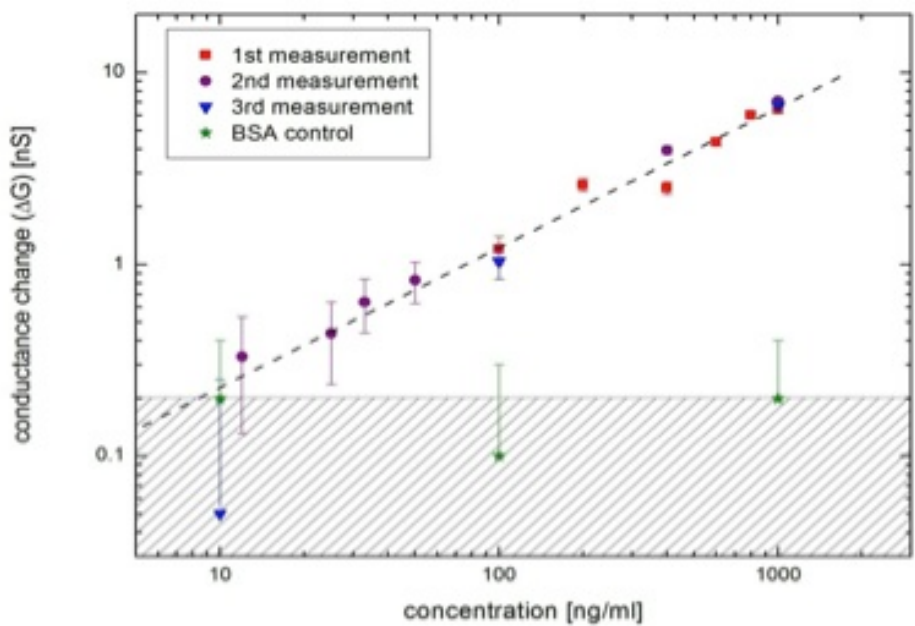

**Figure 2.** Sensing of the melanoma biomarker TROY using FAB-functionalized nanowires. The conductance change of the nanowires in response to different concentrations of TROY in 2 mM PBS buffer (squares, circles and triangles) is compared to the response to BSA (stars) in the same buffer. A linear dependence between conductance change and concentration is observed in the 10 ng/ml to 1 μg/ml regime of TROY concentration.

Among the scalable technologies that can be naturally incorporated into a low-cost handheld configuration for point-of-care use is the silicon-based semiconductor nanosensor. Figure 2 shows that this device has the required sensitivity for TROY detection[19]. The corresponding device, shown in Figure 3, is relatively cheap to manufacture, plus the nanosensor and microfluidics/electronics (Figure 3(c)) in a disposable cartridge configuration are well suited for a POC device. Unlike incumbent technologies, this approach provides immediate read-out and is laboratory/operator independent.

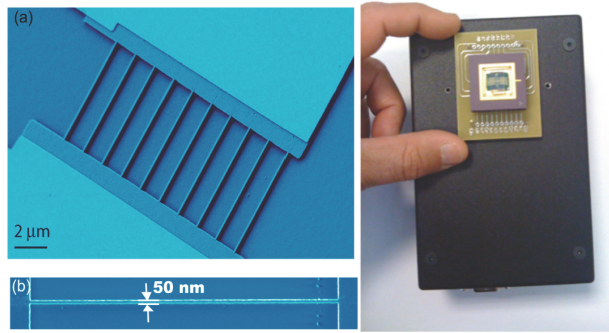

**Figure 3.** (Left) (a) Electrical nanowire sensors fabricated using e-beam lithography; (Right) (b) Single nanowire conductance channel 50 nm in width. (c) Instrument footprint showing chip carrier, and control electronics in the black box.

These silicon nanochannel sensors, which have the ability to detect the presence of a biomarker by directly measuring the change in electric potential upon binding of antigen to an antibody conjugated to the sensor[20,21,22], shown schematically in Figure 4(b) and 4(c), have demonstrated exciting potential as next-generation POC diagnostic devices. The great strides in semiconductor technology allow such sensors to be fabricated cheaply with excellent reproducibility as it has been demonstrated for the detection of breast cancer biomarker CA15.3.[23] Electrical measurements can be configured to provide a simple, effective, and inexpensive method for melanoma biomarker sensing. The combination of state-of-the-art semiconductor nanotechnology and the emergence of a novel and unique melanoma biomarker translates to a first-in-class proof-of-concept diagnostic device.

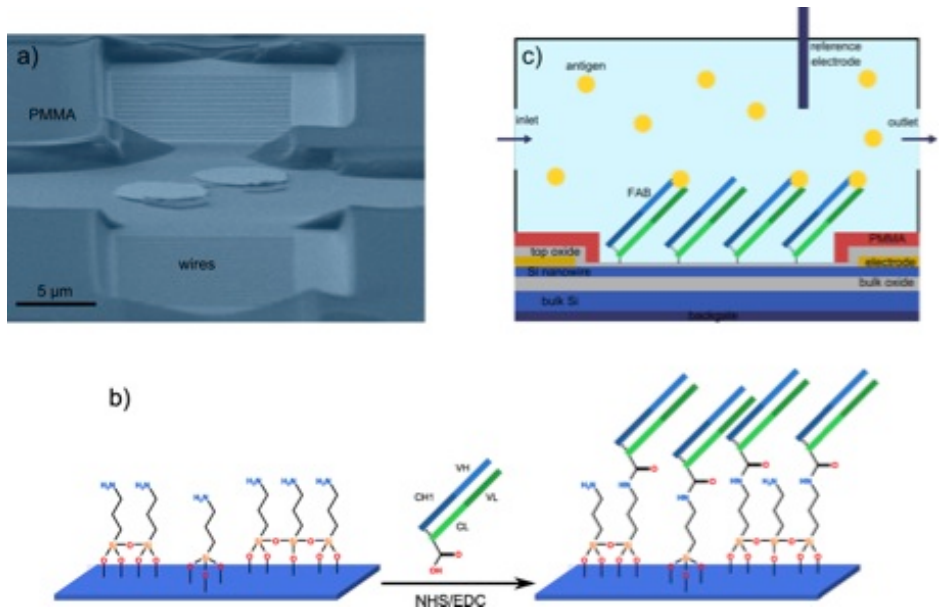

**Figure 4.** (a) Electron micrograph of two separate sets of silicon nanowires with the electrical connections insulated from solution by a micrometer thick layer of poly(methyl methacrylate) (PMMA). (b) Immobilization of the antibody FAB to the silanized nanowire using NHS/EDC coupling chemistry. The FAB has one constant and one variable domain from each heavy and light chain of the antibody (constant heavy (CH), constant light (CL), variable heavy (VH) and variable light (VL). It binds with its C-terminal end to the nanowire exposing the paratope at its variable end to the solution. (c) The sensing nanowires are sealed in a fluid chamber with an inlet and outlet at opposing ends. Their conductance is tunable by a reference electrode immersed in the solution and a back gate connected to the bulk silicon.

## Future outlook

Metastatic melanoma is extremely difficult to treat because tumors are virtually resistant to any type of therapy. Results from the new generation of immunotherapeutic drugs have been impressive but they only appear to be effacious in a subset of patients and only temporarily halt tumor progression.[24] A highly attractive, alternative strategy to reduce melanoma mortality is to determine the presence of melanoma cells before they become clinically detectable by conventional means such as PET-CT, when tumors already are too large and virtually incurable. When tumor cell numbers are low, they may be more sensitive to therapy. Until now, lack of a reliable melanoma biomarker(s) and accompanying ultrasensitive point-of-care diagnostic devices have made testing of this breakthrough concept impossible. However, innovative technologies that build upon decades of advances in semiconductor fabrication and processing can now offer revolutionary potential for changing the standard of care.